\documentclass{article}
\newcommand{\bfr}{\begin{flushright}}
\newcommand{\efr}{\end{flushright}}
 
\begin{document}
\title{High-Energy Scattering by Extreme Dilaton Black Holes
}
\author{Takuya Maki\footnote{maki@jwcpe.ac.jp}
\\
{\small
Japan Women's College of Physical Education, 
Setagaya,~Tokyo 157--8565, Japan
}
\smallskip\\
and
\smallskip\\
Kiyoshi Shiraishi\footnote{shiraish@yamaguchi-u.ac.jp}\\
{\small
Faculty of Science, Yamaguchi University,
Yamaguchi-shi, Yamaguchi 753--8512, Japan}
}
\date{\today
}
\maketitle
\begin{abstract}
We investigate the high and low energy scattering of charged scalar
waves from an extreme dilaton black hole (DBH). The analyses here
correspond to forward scattering processes of two charged scalar
particles with dilaton coupling under a certain extreme condition. We
calculate the scattering amplitude and compare it with the known
results obtained by other methods. We also discuss the case
of incoming wave with a general charge and dilaton coupling.
\\
~
\\
PACS number(s): 04.40.Nr, 04.70.Bw, 11.27.+d.
\end{abstract}

\section{Introduction}
The study of high-energy scattering process including gravity is very
important to explore a possible road to quantum gravity. Since the
seminal papers \cite{thf,amat} appeared, many authors have studied
the scattering of two particles at high energy (for an example, see
\cite{kab}).
The effect of the Coulomb \cite{san,jac,dm3} and dilatonic forces
\cite{dm1,dm2} has been also considered for scattering of charged
particles in the leading order in momentum transfer correction.

The most simple object with mass and charges is a black hole.
Exact solutions for black holes in string-inspired field theory have
been found \cite{GM,GHS,Shi1}.
The interaction of extremely-charged dilatonic
black holes has been studied in the low-velocity limit
 by one of the present authors \cite{Shi2,Shi3}. In particular, the
black hole scattering at a low velocity was eagerly investigated  there.

In the present paper, we investigate the high-energy scattering of a
scalar wave from charged DBHs and reveal the properties by using the
Coulomb wave approximation. The result is approximately equivalent to
the high-energy scattering of two extremely-charged DBHs.
Although our analysis relies on a method of pedagogical fashion,
the results exhibits the interesting result at both low and high energy.

The structure of the present paper is as follows.
In \S\ref{sec2}, we present a system to study by describing the
background fields and wave equations on the background.
In \S\ref{sec3}, we extract the long-range part from the
spherical potential and the scattering of scalar wave by the extreme
DBH is studied. The case with arbitrary charges and dilaton couplings
for the incoming scalar is discussed in \S\ref{sec4}.
The last section (\S\ref{sec5}) is devoted to summary.


\section{The DBH background and a charged scalar field coupled to
dilaton
\label{sec2}}
We suppose an Einstein-Maxwell-dilaton system described by the following
action:
\begin{equation}
S=\int d^{4}x\ \frac{\sqrt{-g}}{16\pi} \left[ R -
2(\nabla \phi)^{2} - e^{-2a\phi} F^{2}
\right] +
\mbox{(surface terms)}\,.
\end{equation}
where $R$ is the scalar curvature, $\phi$ a dilaton,
and $F_{\mu\nu}=\partial_{\mu}A_{\nu}-\partial_{\nu}A_{\mu}$
is the abelian gauge field strength. Here the dilaton coupling $a$ is
considered to take a general value ($a=1$ for effective field
theory of string theory and $a=\sqrt{3}$ for reduced theory from the
Kaluza-Klein compactification).

Let us consider the high-energy scattering of a charged scalar wave
from an extreme DBH in the system.
This process corresponds to the scattering of a charged massive object
from an extreme DBH. In this section, we describe the background
geometry as the target DBH and the wave equation for the
incoming particle.

We assume the metric for an extreme
DBH as \cite{GHS,Shi1}%
\footnote{Strictly speaking, there is a
singularity at $r=0$.
Nevertheless, we use the term ``black hole'' because the extreme case
may still have generic properties of black holes in terms of
their dynamics as a limiting case.
As far as we consider semiclassical scattering cross sections,
the phenomena near the horizon (of which property is sensitive on the
mass-charge relation) can be neglected.}
\begin{equation}
ds^2=-\frac{1}{V^{2/(1+a^2)}}dt^2+V^{2/(1+a^2)}d{\bf x}^2
\label{bg1}
\end{equation}
with
\begin{equation}
V=1+\frac{(1+a^2)M}{r}\,,
\label{bg2}
\end{equation}
where $M$ is the mass of
the DBH. Note that we set the Newton constant $G$ to unity.
The dilaton field configuration and electric potential are given as
\begin{equation}
e^{-2a\phi}=V^{2a^2/(1+a^2)}\,,
\label{bg3}
\end{equation}
and
\begin{equation}
A_t=\frac{1}{\sqrt{1+a^2}}\frac{V-1}{V}\,.
\label{bg4}
\end{equation}
The electric charge $Q$ and the dilatonic charge $\Sigma$ of the DBH 
are considered as
\begin{equation}
Q=\sqrt{1+a^2}M
\end{equation}
and
\begin{equation}
\Sigma=aM\,,
\end{equation}
respectively. These relations are called as an extreme condition among
the mass, charge, and dilatonic charge here.

The classical action for a spinless particle, which has an electric
charge
$q$ and is coupled to the dilaton field with a coupling constant $a'$,
is
\begin{equation}
S_p=-\int ds\left[me^{a'\phi}+qA_\mu\frac{dx^\mu}{ds}\right]\,,
\end{equation}
where $m$ and $q$ are the mass and charge of the particle respectively.
Quantizing the charged scalar particle coupled to the dilaton, we
obtain the scalar wave equation as
\begin{equation}
(D^\mu D_\mu-e^{2a'\phi}m^2)\psi=0\,,
\label{eq8}
\end{equation}
where $\psi$ is the wave function and the covariant derivative is
defined as
\begin{equation}
D_\mu=\partial_\mu+iqA_\mu\,.
\end{equation}

For the background fields of the extreme DBH
(\ref{bg1}),(\ref{bg2}),(\ref{bg3}),(\ref{bg4}), the wave equation
(\ref{eq8}) can be written as
\begin{equation}
{V^{\frac{-2}{1+a^2}}}\nabla^2\psi+V^{\frac{2}{1+a^2}}
\{\omega-q(1+a^2)^{-1/2}(1-V^{-1})\}^2\psi-V^{\frac{-2aa'}{1+a^2}}m^2\psi=0\,,
\label{geq}
\end{equation}
where we assume $\psi\propto e^{-i\omega t}$.

There are particularly simple cases, when the dilaton coupling
constants takes certain values.  

If we assume $a'=a$ and the extreme condition for the
probing wave, 
$q=\sqrt{1+a^2}\, m$, the wave equation becomes
\begin{equation}
\nabla^2\psi+V^{\frac{4}{1+a^2}}\left[
(\omega-m)^2+\frac{2m(\omega-m)}{V}\right]\psi=0\,.
\end{equation}
Some particular cases are
written as follows: When for $a^2=3$, the equation reads
\begin{equation}
\nabla^2\psi+\left[
(\omega-m)^2V+{2m(\omega-m)}\right]\psi=0\,,
\end{equation}
while for $a^2=1$, the equation reads
\begin{equation}
\nabla^2\psi+\left[
(\omega-m)^2V^2+{2m(\omega-m)}V\right]\psi=0\,.
\end{equation}

In the next section, we solve the wave equation to analyze the
scattering of the wave by the extreme DBH.

\section{Long-range effective potential and approximation
\label{sec3}}
In this section, we evaluate the cross section for the wave scattering
in the lowest-order approximation. Here we expand the potential in terms
of inverse of the distance $r$. Expanding the potential in $M/r$ and
dropping terms of higher orders than
$O(r^{-2})$, we get the wave equation
\begin{equation}
\left[\nabla^2+k^2+2\alpha\left(\frac{M}{r}\right)+
\beta\left(\frac{M}{r}\right)^2\right]\psi=0\,,
\end{equation}
where $\alpha$ and $\beta$ are given by
\begin{eqnarray}
\alpha&=&(\omega-m)\{2\omega+(1-a^2)m\}=\frac{\omega-m}{M}\{s-qQ-(M-m)^2\}\nonumber
\\
&=&2k^2-(1+a^2)m(\sqrt{k^2+m^2}-m)\,,\nonumber \\
\beta&=&2(3-a^2)(\omega-m)(\omega-ma^2)\nonumber \\
&=&2(3-a^2)[k^2-(1+a^2)m(\sqrt{k^2+m^2}-m)]\,.
\end{eqnarray}
Here, we use $k=\sqrt{\omega^2-m^2}$. The Mandelstam variables $s$ and
$t$ are defined by:
\begin{eqnarray}
s&=&2M\omega+M^2+m^2\,,\\
t&=&-|{\bf k}-{\bf k}'|^2=-(2k\sin(\theta/2))^2
\,,
\end{eqnarray}
where $\theta$ is the scattering angle.

We take the lowest order approximation, where
the term of $O(M^2/r^2)$ is omitted. Then, the wave equation reduces to
\begin{equation}
\left(\nabla^2+k^2+\frac{2\alpha M}{r}\right)\psi=0\,.
\label{Coul}
\end{equation}
The asymptotic form of the solution for scattering is \cite{kab,dm3}
\begin{equation}
\psi\approx e^{i[kz-\frac{\alpha
M}{k}\ln k(r-z)]}+\frac{e^{i(kr+\frac{\alpha M}{k}\ln
2kr)}}{r}f(\theta)\,,
\end{equation}
and then the scattering amplitude $f(\theta)$ is given as \cite{kab,dm3}
\begin{eqnarray}
f(\theta)&=&\frac{\alpha M}{2k^2\sin^2(\theta/2)}\exp\left[
i\frac{\alpha M}{k}\ln(\sin^2 \theta/2)+2i\delta_0\right]\nonumber \\
&=&\frac{1}{2ik}\frac{\Gamma(1-i\alpha M/k)}{\Gamma(i\alpha M/k)}
\left(\frac{4k^2}{-t}\right)^{1-i\alpha M/k}\,,
\end{eqnarray}
where the phase shift is obtained as
\begin{equation}
\delta_0=\arg\Gamma(1-i{\alpha M}/{k})\,.
\end{equation}

The scattering cross section of the lowest-order, in other words, in the
Coulomb wave approximation is
\begin{eqnarray}
\frac{d\sigma}{d\Omega}&=&|f(\theta)|^2\nonumber \\
&=&\frac{1}{4k^2\sin^4(\theta/2)}\left|\frac{\Gamma(1-i\alpha M/k)}{
\Gamma(i\alpha
M/k)}\right|^2=\frac{\alpha^2M^2}{4k^4\sin^4(\theta/2)}\,.
\label{cs}
\end{eqnarray}

For the case that $k\ll m$, we find
\begin{equation}
\frac{d\sigma}{d\Omega}=\frac{(3-a^2)^2M^2}{16\sin^4(\theta/2)}\,.
\label{tr}
\end{equation}

This result agrees with the previous result (of one of the
present authors) obtained by the classical analysis of the moduli space
of slowly-moving extreme DBHs
\cite{Shi3}.  For
$a=1$, the expression (\ref{tr}) coincides with Eq.~(3.14) in
Ref.~\cite{Shi3}. For $a^2=3$, almost no scattering occurs at the low
energy, since there is no force between two extremal DBHs in the
slowly-moving limit.

On the other hand, in the high-energy limit, $\alpha M/k\approx
s-Qq-(M-m)^2$. Thus
\begin{eqnarray}
\frac{d\sigma}{d\Omega}
&=&\frac{[s-Qq-(M-m)^2]^2}{4k^2\sin^4(\theta/2)}\,.
\label{cs25}
\end{eqnarray}

Then the result (\ref{cs25}) agrees with one of 't Hooft
\cite{thf} when $M-m$ is neglected. The gravitational interaction is
certainly dominant in the Planckian energy scale.
It should be noted that no effect of  dilaton couplings and charges
can be seen either in the lowest order approximation in high-energy
scattering.

The validity of the approximation is guaranteed by assuming the large
impact parameter, because only the lowest order in $M/r$ is expected
to be dominant. This is equivalent to the case with small momentum
transfer, or nearly forward scattering. 

As further exploration on the validity of the approximation,
we will estimate the feasible value for the impact parameter.
If we expand the wave function as the angular momentum eigenvalue $l$,
we find the centrifugal potential as $l(l+1)/r^2$. Thus, the 
$O(M^2/r^2)$ term of potential can be neglected if
\begin{equation}
l(l+1)\gg |\beta| M^2\,.
\end{equation}

Since the impact parameter is estimated semiclassically as
$b\approx\sqrt{l(l+1)}/k$, 
\begin{equation}
b\gg \sqrt{|\beta|} M/k\,.
\end{equation}

For the case that $k\ll m$, the approximation holds for
\begin{equation}
b\gg \sqrt{|(3-a^2)(1-a^2)|}M\,,
\end{equation}
while  the approximation holds for the case that $k\gg m$
\begin{equation}
b\gg \sqrt{2|(3-a^2)|}M\,.
\end{equation}

\section{The case with arbitrary charges and dilaton couplings for the
incident wave
\label{sec4}}

The wave equation with an arbitrary coupling and charge (\ref{geq})
is approximated to be in the form of Eq. (\ref{Coul}). Then the
coefficient $\alpha$ is
\begin{eqnarray}
\alpha M&=&M[(2\omega-\sqrt{1+a^2}q)\omega+(aa'-1)m^2]\nonumber \\
&=&(2\omega^2-m^2)M-Qq\omega+\Sigma\Sigma'm\,,
\end{eqnarray}
where $\Sigma'=a'm$.

Hence, for arbitrary couplings, when the extreme condition is not
satisfied, the scattering of very low energy is nothing but the
Rutherford scattering with the cross section
\begin{equation}
\frac{d\sigma}{d\Omega}
=\frac{\alpha^2M^2}{4k^4\sin^4(\theta/2)}=\frac{(Mm-Qq+\Sigma\Sigma')^2m^2}{4k^4\sin^4(\theta/2)}\,,
\end{equation}
for $Mm-Qq+\Sigma\Sigma'\ne 0$. This result is trivial because of the
existence of residual long-range static forces. Of course, at very high
energy, the cross section is similar to (\ref{cs25}) because of the
gravity dominance.

\section{Summary and outlook
\label{sec5}}
We have investigated the scattering of charged scalar wave
from an extreme DBH. The scalar field is supposed to satisfy the
extreme condition. This corresponds to the Planckian forward scattering
process of two charged scalar particles with the extreme condition. We
have evaluated the scattering cross section for the corresponding case
of a large impact parameter.

We have found that our simple analyses can reveal the behavior of the
eikonal scattering from black holes. Therefore we
intend to study the scattering with excited states such as in the
Kaluza-Klein and string theory in the similar manner and compare the
results with ones obtained from various methods.





\bibliographystyle{apsrev4-1}

\begin{thebibliography}{99}

\bibitem{thf} 't Hooft G. Graviton dominance in ultra-high-energy
scattering. Phys. Lett. {\bf B} 1987; 198: 61-3; On the factorization of
universal poles in a theory of gravitating point particles. Nucl. Phys.
{\bf B} 1988; 304: 867-76.

\bibitem{amat} Amati D, Ciafaloni M, Veneziano G. Superstring
collisions at planckian energies. Phys. Lett. {\bf B} 1987; 197: 81-8;
Classical and quantum gravity effects from planckian energy
superstring collisions. Int. J. Mod. Phys. {\bf A} 1988; 3: 1615-61;
Higher-order gravitational deflection and soft bremsstrahlung in
planckian energy superstring collisions. Nucl. Phys. {\bf B} 1990; 347:
550-80.


\bibitem{kab} Kabat D, Ortiz M. 
Eikonal quantum gravity and planckian scattering.
Nucl. Phys. {\bf B} 1992; 388: 570-92.

\bibitem{san} Loust\'o C, S\'anchez N. 
The curved shock wave space-time of ultrarelativistic charged particles
and their scattering.
Int. J. Mod. Phys. {\bf A}
1990: 5: 915-38.

\bibitem{jac} Jackiw R, Kabat D, Ortiz M. 
Electromagnetic fields of a massless particle and the eikonal.
Phys. Lett. {\bf B} 1992; 277: 148-52.

\bibitem{dm3} Das S, Majumdar P. 
Aspects of Planckian scattering beyound the eikonal
Pramana 1998; 51: 413-9.

\bibitem{dm1} Das S, Majumdar P. 
Shock wave mixing in Einstein and dilaton gravity.
Phys. Lett. {\bf B} 1995; 348: 349-54.

\bibitem{dm2} Das S, Majumdar P.
Eikonal particle scattering and dilaton gravity.
Phys. Rev. {\bf D} 1997; 55: 2090-8.

\bibitem{GM}
Gibbons GW, Maeda K.
Black holes and membranes in higher-dimensional theories with dilaton
fields.
Nucl. Phys. {\bf B} 1988; 298: 741-75.
\bibitem{GHS}
Garfinkle D, Horowitz G, Strominger A.
Charged black holes in string theory.
Phys. Rev. {\bf D} 1991; 43: 3140-3; (Erratum) 1992; 45: 3888.

\bibitem{Shi1} Shiraishi K.
Multi-Centered Solution for Maximally-Charged Dilaton Black holes in
Arbitrary Dimensions.
J. Math. Phys. 1993; 34: 1480-6.

\bibitem{Shi2} Shiraishi K.
Moduli Space Metric for Maximally Charged Dilaton Black holes.
Nucl. Phys. {\bf B} 1993; 402; 399-410.

\bibitem{Shi3} Shiraishi K.
Classical and Quantum Scattering of
Maximally Charged Dilaton Black holes.
Int. J. Mod. Phys. {\bf D} 1993; 2: 59-77.








\end{thebibliography}


\end{document}